\documentclass[amsmath,amssymb,amsfonts,aps,pra,preprint,groupedaddress,showpacs]{revtex4-1}
\usepackage{graphicx}
\usepackage{dcolumn}
\usepackage{bm}
\usepackage{color}

\usepackage{amsmath}

\usepackage{color}
\usepackage{amssymb}
\usepackage{hyperref}

\newcommand{\ud}{\mathrm{d}}
\newcommand{\ui}{\mathrm{i}}
\newcommand{\be}{\begin{equation}}
\newcommand{\ee}{\end{equation}}
\newcommand{\bea}{\setlength\arraycolsep{1pt}\begin{eqnarray}}
\newcommand{\eea}{\end{eqnarray}}
\newcommand{\br}{{\bf r}}
\newcommand{\bu}{{\bf u}}

\begin{document}

\title{Spherical fields as nonparaxial accelerating waves}  
\author{Miguel A. Alonso$^{1,*}$ and Miguel A. Bandres$^2$}

\affiliation{
$^1$The Institute of Optics, University of Rochester, Rochester, NY 14627, USA
\\
$^2$Instituto Nacional de Astrof\'isica, \'Optica y Electr\'onica\\
Calle Luis Enrique Erro No. 1, Sta. Ma. Tonantzintla, Pue. CP 72840, M\'exico\\
$^*$Corresponding author: alonso@optics.rochester.edu
}

\begin{abstract}
We introduce nonparaxial spatially accelerating waves 
whose two-dimensional transverse profiles propagate along semicircular trajectories while approximately preserving their shape. We derive these waves by considering imaginary displacements on spherical fields, leading to simple closed-form expressions. The structure of these waves also allows the closed-form description of pulses.\end{abstract}

\maketitle

The so-called ``accelerating waves'' have received considerable attention due to their remarkable properties: they preserve their intensity profile under propagation, but the profile's features follow a curved path. This behavior seems at odds with Ehrenfest's theorem, which in this context states that the transverse intensity centroid of any free paraxial beam follows a straight path. The best-known example is that of Airy beams \cite{BerryBalazs,Dogariu,Vo}, although more general families of accelerating paraxial beams exist 
\cite{ApA,AcB}. 

The strange behavior of these beams can be understood through two observations. First, their transverse intensity is not integrable, so they require infinite power 
and their transverse centroid is not well defined (i.e., Ehrenfest's theorem is not violated). Second, they are associated with rays forming parabolic caustics, so the transverse intensity maximum is due to different ray bundles at different propagation distances. That is, one should not think of this maximum as a ``physical entity'' following a curved path. 
In real optical systems, the plane-wave spectrum is limited, leading to finite-power approximations to accelerating beams whose intensity profile is roughly preserved over some distance. While the intensity maximum of these beams still follows a curved path within this range, the now well-defined intensity centroid describes a straight line.

Nonparaxial accelerating waves in two dimensions,
 whose intensity profiles are roughly preserved within a range of propagation distances, and whose maxima follow semicircular paths, were recently proposed \cite{Kaminer,Zhang}. These waves result from truncating the backward-propagating components of fields with cylindrical symmetry. Monochromatic and pulsed waves following other nonparaxial non-circular paths have also been implemented \cite{Froehly,Courvoisier}.

In this Letter, we extend these ideas to three-dimensional waves by starting with fields with rotational symmetry. 
Our solutions have a two-dimensional transverse intensity profile that is approximately invariant under propagation and whose maxima follow circular paths. To construct these waves we apply the concept of imaginary displacements \cite{BerryComplexShift}. By this means, unlike the solutions generated by imposing an abrupt exit pupil \cite{Kaminer,Zhang}, we are able to give simple close-form solutions.

\begin{figure}[htp]
\centerline{\includegraphics[width=12cm]{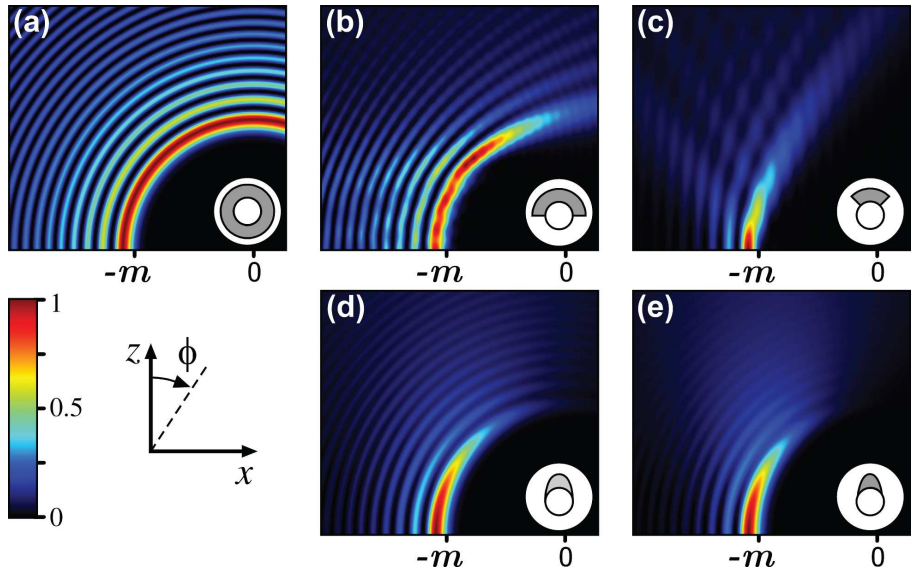}}
\caption{\label{FigBess} (Color online) Intensities over a $90\times80$ rectangle (in units of $k^{-1}$) of $m=40$ Bessel fields: (a) unapertured, apertured with (b) $\alpha=\pi$ and (c) $\alpha=\pi/2$, and apodized with (d) $q=1.88$ and (e) $q=3$. The insets indicate the power angular spectrum.} 
\end{figure}

We begin by considering two-dimensional waves, i.e., solutions to the two-dimensional Helmholtz equation. One such solution is the Bessel field \cite{Uniform},
\bea
\psi_m(x,z)&=&u_0\ui^mJ_m\left[k\sqrt{x^2+z^2}\right]\exp\left[\ui m\arctan(z,x)\right],
\eea
where $u_0$ is a constant, $m$ is an integer (assumed here to be positive), $J_m$ is a Bessel function of the first kind, $k$ is the wavenumber,
 $(x,z)$ are 
Cartesian coordinates where $z$ is regarded as the propagation direction, and $\arctan(\cos\phi,\sin\phi)=\phi$. This field has an intensity profile that is exactly preserved not over lines of constant $z$ but over lines crossing the origin, and its maximum follows a circular caustic of radius $m/k$. A section of this  profile is shown in Fig.~\ref{FigBess}(a)
. For large $m$, the radial profile of this field approaches the profile of a paraxial Airy beam over a finite radial region near $m/k$. In fact, accurate approximations to these fields can be written in terms of Airy functions even for modest values of $m$ \cite{Uniform}.

Notice, though, that Bessel fields are composed of plane waves traveling in all directions:
\be
\psi_m(x,z)=\int_{-\pi}^\pi A_m(\phi)\,\exp[\ui k(x\sin\phi+z\cos\phi)]\,\ud\phi,\label{BesselAS}
\ee
where $\phi$ is the plane waves' propagation angle from the $z$ axis, and $A_m(\phi)=u_0\exp(\ui m\phi)/2\pi$ is the angular spectrum. 
In practice, it is difficult to generate fields whose plane-wave spectrum extends over all directions, so fields with spectra confined to forward propagation, i.e., $A\left(\phi\right)=0$ for $\cos\phi<1$, are usually considered.
For this purpose, Kaminer {\it et al.} \cite{Kaminer} proposed limiting the integration in Eq.~(\ref{BesselAS}) to the forward semicircle $\phi\in[-\pi/2,\pi/2]$, which results in a forward-propagating wave with accelerating characteristics. Smaller ranges $\phi\in[-\alpha/2,\alpha/2]$ for $0<\alpha\le\pi$ can also be used (and would be easier to implement in practice). The intensities of these fields for $\alpha=\pi$ and $\pi/2$, are shown in Figs.~\ref{FigBess}(b), \ref{FigBess}(c), respectively. Notice that the intensity maximum propagates approximately along a circular path until its angle with respect to $z$ is slightly below $\alpha/2$. However, for any $\alpha\neq2\pi$, the fields must be evaluated numerically, and as shown in Fig.~\ref{FigBess}(b)-\ref{FigBess}(c), the hard integration limits introduce appreciable rippling in the intensity. 

To avoid these issues, we propose to use instead a Gaussian-like apodization factor $\exp[kq(\cos\phi-1)]$ in the angular spectrum. This is formally equivalent to multiplying by $\exp(-q)$ and performing an imaginary shift $z\to z-\ui q/k$ \cite{BerryComplexShift}. The resulting field is given simply by 
\be
\psi_{m}(x,z;q)=\exp(-q)\,\psi_m(x,z-\ui q/k).
\ee
The intensities of these fields are shown in Figs.~\ref{FigBess}(d)-\ref{FigBess}(e) for $q=1.88$ and $3$. Note that the intensity maximum decreases smoothly away from $z=0$, and follows a circular path until this path forms an angle of about $2\arccos[I_1(q)/I_0(q)]$ with respect to 
$z$, with $I_n$ being the modified Bessel function of the first kind. 
While these fields contain some counter-propagating components, these represent only a fraction $[1-L_0(2q)/I_0(2q)]/2$ of the total power, where $L_n$ is the modified Struve function of the first kind. For the cases in Figs.~\ref{FigBess}(d)-\ref{FigBess}(e), this power fraction is about $1\%$ and $0.082\%$, respectively.

\begin{figure*}[htp]
\begin{center}
\includegraphics[width=6.4in]{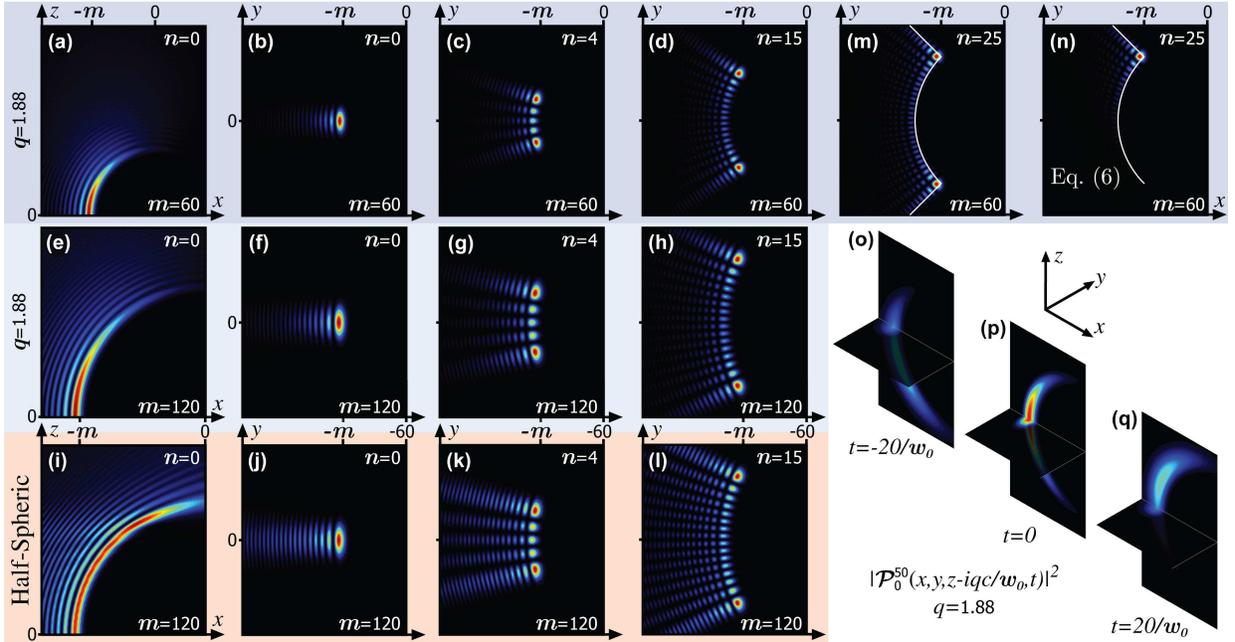}
\end{center}
\caption{\label{FigAHG}(Color online) Intensities over sections of size $180\times156$ (in units of $k^{-1}$) of the $z=0$ and $x=0$ planes for (a-h,m) the waves in Eq.~(\ref{UM}), (i-l) truncated spherical waves with $\alpha=\pi$, (n) the waves in Eq.~(\ref{UP}), and (o-q) the pulses in Eq.~(\ref{pls}) (movie online). The cross-section of the caustic is shown in (m) and (n).} 
\end{figure*}

We extend these ideas to create three-dimensional accelerating waves by using multipoles (separable in spherical coordinates), whose plane-wave spectra are spherical harmonics $Y_{l,m}$, aligned with the $y$ axis for convenience:
\bea
\Lambda^m_{l}(\br)&=&\int Y^m_{l}(\theta,\phi)\exp(\ui k\br\cdot\bu)\,\ud\Omega\nonumber\\&=&4\pi\ui^l j_l(k\sqrt{x^2+y^2+z^2})\nonumber\\
&\times&Y^m_{l}\left[\arctan\left(y,\sqrt{x^2+z^2}\right),\arctan(z,x)\right],\label{multipole}
\eea
where $\br=(x,y,z)$, $j_l$ is a spherical Bessel function of the first kind, and the integral is over the unit vector $\bu=(\sin\theta\sin\phi,\cos\theta,\sin\theta\cos\phi)$. 
We then define spherical waves with imaginary displacements as
\be
\Psi^m_n(\br;q)=u_0\exp(-q)\Lambda^m_{m+n}\left(x,y,z-\ui q/k\right),\label{UM}
\ee
for $n\ge0$. For $q=0$ (no apodization) these fields have exact rotational symmetry about the $y$ axis. For $q\tilde{>}2$, on the other hand, the counter-propagating components are largely suppressed, and for $m\gg n$, these fields can be regarded as nonparaxial versions of Airy-Hermite-Gaussian beams, whose amplitude profile in $y$ is a Hermite-Gauss function of order $n$ (i.e., $n$ specifies the number of zeroes in the $y$ direction), while in the $xz$ plane they behave as Airy beams. The intensity of some of these fields is shown in Figs.~\ref{FigAHG}(a)-\ref{FigAHG}(h).
Of course, instead of apodizing through an imaginary displacement, one can define apertured versions where the integral in Eq.~(\ref{multipole}) is limited to a forward-propagating section of the sphere [$\sin\theta\cos\phi\ge\cos(\alpha/2)$] and must be evaluated numerically. The case of half-spherical waves ($\alpha=\pi$) is shown in Figs.~\ref{FigAHG}(i)-\ref{FigAHG}(l). Whether apodized or apertured, these spherical waves present an intensity maximum (or maxima for $n>0$) that follows a circular path of radius slightly larger than $m/k$, cf. Figs.~\ref{FigAHG}(a,e,i). 

It is seen from Fig.~\ref{FigAHG} that, for larger $n$, the field's transverse intensity profile is noticeably curved. This is because the caustic, whose cross-section is shown as a white line in Fig.~\ref{FigAHG}(m), is composed of segments of a sphere of radius $(m+n)/k$ and a (two-sided) cone of half-angle $\arcsin[m/(m+n)]$ around the $y$ axis, both centered at the origin. The intensity then presents two cusps, one at each intersection of the sphere and the cone, and therefore resembles two joined Airy beams. One of the cusps can be suppressed, e.g., by combining three of these fields as
\be
\Psi^m_n
-\frac{\ui}2\left(\Psi^m_{n+1}-\Psi^m_{n-1}\right),\label{UP}
\ee
The intensity of this field over the $z=0$ plane is shown in Fig.~\ref{FigAHG}(n). Note that it indeed resembles a single Airy beam whose cusp at the $z=0$ plane is approximately at $y=\sqrt{(2m+n)n},\,x=-m$, but where one of the caustic sheets is noticeably curved. For $n\approx(\sqrt{2}-1)m\gg1$, this curvature becomes less appreciable and the field increasingly resembles an Airy beam.

The fields described so far are monochromatic. We now define closed-form pulsed solutions to the wave equation that follow similar paths. This is possible because in Eq.~(\ref{multipole}), $k$ appears only as a factor in the argument of 
$j_l$, which can be written as a finite sum of powers times exponentials of its argument:
\be
j_l(u)=
\sum_{p=0}^l \frac{(2l-p)!\,[(-\ui u)^p\exp(\ui u)-(\ui u)^p\exp(-\ui u)]}{2^{l-p+1}\,\ui\,u^{l+1}\,p!\,(l-p)!}.\label{SphBess}
\ee
Substituting $u=\omega r/c$ (where $\omega$ is the frequency and $c$ the speed of light) in Eq.~(\ref{SphBess}) and multiplying by $\omega^{l+1}$ leads to an expression where each term includes $\omega$ as a non-negative power times an exponential. By using this result together with Eq.~(\ref{multipole}) and applying standard Fourier relations, we define the following pulsed solutions 
\bea
P^m_n(\br,t)&=&\int\tilde{f}(\omega)\omega^{m+n+1}\Lambda_{m+n}^m(\br)\exp(-\ui\omega t)\,\ud\omega\nonumber\\
&=&4\pi\,Y_{m+n}^m(\theta,\phi)\left(\frac{\ui c}{2r}\right)^{m+n+1}\nonumber\\
&\times&\sum_{p=0}^{m+n}\frac{(2m+2n-p)!}{\,p!\,(m+n-p)!}\left(\frac{2r}c\right)^p\nonumber\\
&\times&\left[(-1)^pf^{(p)}\left(t+\frac{r}{c}\right)-f^{(p)}\left(t-\frac{r}{c}\right)\right],
\label{pls}
\eea
where $\tilde{f}(\omega)$ is an arbitrary spectrum, $f(t)$ is its inverse Fourier transform, $f^{(p)}$ is the $p$th derivative of $f$, and $(r,\theta,\phi)$ are spherical coordinates with respect to the $y$ axis. Figures~\ref{FigAHG}(o)-\ref{FigAHG}(q) show cross sections at three times for such a pulse for $f(t)=\exp(-t^2\Delta^2/2-\ui\omega_0t)$ [for which $f^{(p)}$ can be found in closed form in terms of Hermite polynomials] with $\Delta=\omega_0/10$, $m=50$, $n=0$, and using an imaginary displacement in $z$ of size $1.88c/\omega_0$. As in the monochromatic case, for larger values of $n$ these pulses display two cusps, one of which can be eliminated through the combination $P^m_n-(P^m_{n+1}+P^m_{n-1})/2$. It is worth noting that, unlike the pulses in \cite{Courvoisier}, these pulses do not run along the caustic, but rather inhabit it simultaneously at a given range of times.

In conclusion, we proposed nonparaxial accelerating fields given by simple closed-form expressions. While these fields are scalar, polarization can be easily incorporated by either using vector multipoles or applying suitable operators. Also, the paths followed by the maxima are circular due to the use of solutions separable in spherical coordinates. Using other coordinate systems would result in other shapes of the caustic sheets. 

MAA acknowledges support from the National Science Foundation (PHY-1068325).


\begin{thebibliography}{10}
\newcommand{\enquote}[1]{``#1''}

\bibitem{BerryBalazs}
M.~V. Berry and N.~L. Balazs, Am. J. Phys. \textbf{47}, 264 (1979).

\bibitem{Dogariu}
G.~A. Siviloglou, J.~Broky, A.~Dogariu, and D.~N. Christodoulides, Phys. Rev.
  Lett. \textbf{99}, 213901 (2007).

\bibitem{Vo}
S.~Vo, K.~Fuerschbach, K.~P. Thompson, M.~A. Alonso, and J.~P. Rolland, J. Opt.
  Soc. Am. A \textbf{27}, 2574 (2010).

\bibitem{ApA}
M.~A. Bandres, Opt. Lett. \textbf{33}, 1678 (2008).

\bibitem{AcB}
M.~A. Bandres, Opt. Lett. \textbf{34}, 3791 (2009).

\bibitem{Kaminer}
I.~Kaminer, R.~Bekenstein, J.~Nemirovsky, and M.~Segev, Phys. Rev. Lett.
  \textbf{108}, 163901 (2012).

\bibitem{Zhang}
P.~Zhang, Y.~Hu, D.~Cannan, A.~Salandrino, T.~Li, R.~Morandotti, X.~Zhang, and
  Z.~Chen, Opt. Lett. \textbf{37}, 2820 (2012).

\bibitem{Froehly}
L.~Froehly, F.~Courvoisier, A.~Mathis, M.~Jacquot, L.~Furfaro, R.~Giust, P.~A.
  Lacourt, and J.~M. Dudley, Opt. Express \textbf{19}, 16455 (2011).

\bibitem{Courvoisier}
F.~Courvoisier, A.~Mathis, L.~Froehly, R.~Giust, L.~Furfaro, P.~A. Lacourt,
  M.~Jacquot, and J.~M. Dudley, Opt. Lett. \textbf{37}, 1736 (2012).

\bibitem{BerryComplexShift}
M.~V. Berry, J. Phys, A: Math. Gen. \textbf{27}, L391 (1994).

\bibitem{Uniform}
M.~V. Berry, Sci. Prog., Oxf. \textbf{57}, 43 (1969).

\end{thebibliography}
\end{document}